# Bipolaron Model of Superconductivity in Chalcogenide Glasses

Liang-You Zheng, Bo-Cheng Wang[a], Shan T. Lai[b,c]

Center for Molecular Dynamics and Energy Transfer
Department of Chemistry
The Catholic University of America, Washington, DC 20064, USA

*a) Department of Chemistry, Tamkang University, 151 Ying-chuan Road Tamsui, Taipei County Taiwan 25137, ROC*
b) Vitreous State Laboratory, the Catholic University of America, Washington, DC 20064, USA
c) To whom requests for reprints should be addressed

*Abstract* : In this paper we propose a small bipolaron model for the superconductivity in the Chalcogenide glasses (c-As$_2$Te$_3$ and c-GeTe). The results are agree with the experiments.

# 1. Introduction

Chalcogenide glasses are very good glasses for the optical materials [1]. However, some of them have superconductivity behaviors. For example, As$_2$Te$_3$ and c-GeTe [2]. Based on M. Foygel et al work, in the present paper we propose a two site small bipolaron model[3] as a superconductivity mechanism. In that case, the bipolaron model of vibonic coupling ( taking the long wave length approximation ) we have proposed either can apply to high temperature superconductivity [4] or room temperature superconductivity [5]. In this paper low temperature superconductivity either, just only we hold the parameter $\mu$ -- chemical potential of boson ( bipolaron ) or we have $m**$ --the mass of bipolaron and $n$ --the concentration of bipolaron.

# 2. Model Hamiltonian

From the book of N.F. Mott and Davis E.A. [6] we can see that the structure of crystalline c-As$_2$S$_3$ as follows ( As$_2$Te$_3$ should be the same )

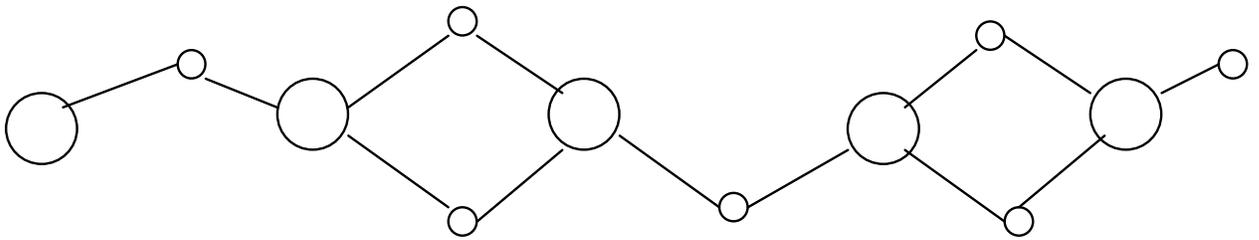

Fig. 1 Structure of Crystalline As$_2$S$_3$
Big circle is As-atom, small circle is S-atom

When the atomic sites of crystal vibrate longitudinally the acoustic wave propagates along the longitudinal direction. Here we regard the four atoms( two As atoms and two S atoms ) as a cluster, an entity. So we can call our model as two site small bipolaron model in one dimension.

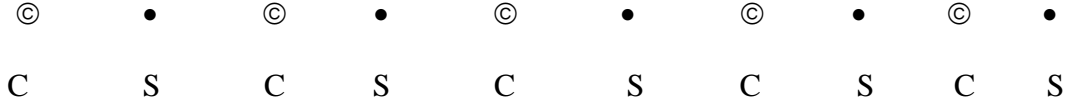

Fig. 2　C----represents the cluster ( two As atoms and two S atoms )
　　　　S----represents the Sulfur atom

In general, the Hamiltonian of the electronic and vibrational system may be written as

$$H = H_e + H_a, \qquad (1)$$

where

$$H_e = \sum_i h_i = \sum_i \left[\frac{p_i^2}{2m} + \sum_n V(r_i - R_n)\right], \qquad (2)$$

$$H_a = \sum_n \left[\frac{K}{2}(u_{n+1} - u_n)^2 + \frac{M_n}{2}\dot{u}_n^2\right]. \qquad (3)$$

In the tight-binding approximation the electronic Hamiltonian reads as

$$H_e = -\sum_n t(R_{n+1} - R_n)(C_{n+1}^+ C_n + C_n^+ C_{n+1}) \qquad (4)$$

where $t(R_{n+1} - R_n)$ is the interaction of two nearest neighbor ions with the the electron. $R_{n+1}$ and $R_n$ are the instantaneous position of the $(n+1)$th ion and the $n$th ion. Because the displacement of the $n$th ion $u_n$ around the instantaneous position $R_n$ is very small, the difference between the distance of neighboring ions reads as $R_{n+1} - R_n = R_{n+1}^{(0)} - R_n^{(0)} + (u_{n+1} - u_n)$ and the distance of equilibrium position $R_{n+1}^{(0)} - R_n^{(0)} = a$, the lattice constant, is very small. We have $u_{n+1} - u_n \ll a$. Therefore, the interaction can be expanded as

$$t(R_{n+1} - R_n) = t_0 - \gamma(u_{n+1} - u_n) \qquad (5)$$

where $t_0 = t(R_{n+1}^{(0)} - R_n^{(0)})$ is the interaction between the electron and the nearest-neighbor ions in their equilibrium position and $\gamma = dt/dx$ is the rate of the change of the interaction with respect to the distance between ions within a unit cell. Substituting Eq.(5) into Eq.(4) and taking into account the spin of electron, we have

$$H_e = -\sum_{n,s}[t_0 - \gamma(u_{n+1} - u_n)](C^+_{n+1,s}C_{n,s} + C^+_{n,s}C_{n+1,s}) \tag{6}$$

The total Hamiltonian is now written as the summation of eq.(3) and eq.(6)

$$H = H_e + H_a = -\sum_{n,s}[t_0 - \gamma(u_{n+1} - u_n)](C^+_{n+1,s}C_{n,s} + C^+_{n,s}C_{n+1,s})$$

$$+ \sum_n \left[\frac{K}{2}(u_{n+1} - u_n)^2 + \frac{M_n}{2}\dot{u}_n^2\right] \tag{7}$$

We can prove that Eq.(6) is equivalent to Eq.(2) in the tight-binding approximation. Su et al.[7,8] first applied Eq.(7) to the trans-polyacetylene. They considered eq.(5) as the standard form of the electron-phonon coupling in the metals.

## 3. The Canonical Transformation

For simple, we treat the singlet bipolaron. In that case, from the total system we treat the system of no spin. For the convenience of illustration, we drop the spin of electron. The Hamiltonian of the electron is written as

$$H_e = -\sum_n [t_0 - \gamma(u_{n+1} - u_n)](C^+_{n+1}C_n + C^+_n C_{n+1}) \tag{8}$$

Evaluating the spectrum of the electron is equivalent to diagonalization the Hamiltonian operator. For this, we transform the operators $C_n$ and $C^+_n$ from position space to momentum space.

$$C_{n_0} = \frac{1}{\sqrt{N}}\sum_k \exp(-i2\pi k n_0 a)(C^e_k + C^0_k)$$

$$C_{n_e} = \frac{1}{\sqrt{N}}\sum_k \exp(-i2\pi k n_e a)(C^e_k - C^0_k) \tag{9}$$

where $C_{n_o}$ is the operator of an electron on the site of odd-numbered ion. $C_{n_e}$ is the operator of an electron on the site of even-numbered ion.

For the displacement of crystal lattice, we have

$$u_n = \sum_q \sqrt{\frac{\hbar}{2NM\omega_q}} \left(a_q + a_{-q}^+\right) \exp(-i2\pi qna) \quad (10)$$

where $a_q$ and $a_q^+$ are phonon operators and $\omega_q$ is the frequency of the phonon. We obtain the Hamiltonian of an electron as follows

$$H_e = \sum_k [E_0(k)(C_k^{e+}C_k^e + C_k^{0+}C_k^0) - \sum_q B_{k,q}(a_q + a_{-q}^+)(C_k^{e+}C_k^e + C_k^{0+}C_k^0) \quad (11)$$

where

$$E_0(k) = -2t_o \cos(2\pi ka) \quad (12)$$

and

$$B_{k,q} = 2\gamma \cos(2\pi ka) \sqrt{\frac{\hbar}{2NM\omega_q}} [i2\pi qa] \quad (13)$$

are under the limit of long wave length.

If we assume that the probability of an electron creation or annihilation, respectively, around the odd-numbered ions is equivalent to the probability of creation and annihilation of an electron around even-numbered ions. For two site small bipolaron that is true. i.e.

$$C_k^{e+}C_k^e = C_k^{o+}C_k^o \quad \text{and} \quad C_k^{e+}C_k^e + C_k^{o+}C_k^o = n_k^e + n_k^o = B_k^+ B_k \quad (14)$$

Therefore we have

$$H_e = H_{e-B} = \sum_k [E_0(k) - H_k] B_k^+ B_k. \quad (15)$$

Here

$$H_k = \sum_q B_{k,q}(a_q + a_{-q}^+) \quad (16)$$

The total Hamiltonian is now rewritten as

$$H = \sum_k [E_0(k) - H_k] B_k^+ B_k + \sum_q \hbar\omega_q \left(a_q^+ a_q + \frac{1}{2}\right) \quad (17)$$

Now let us make the canonical transformation as follows[9]

$$H_T = \exp(-s) H \exp(s) \qquad (18)$$

where

$$s = \sum_k \sum_q \frac{B_{k,q}}{\hbar \omega_q} \left( a_q + a_{-q}^+ \right) B_k^+ B_k \qquad (19)$$

Therefore, we obtain

$$H_T = \sum_k \left[ E_0(k) - \Delta_k \right] B_k^+ B_k + \sum_q \hbar \omega_q \left( a_q^+ a_q + \frac{1}{2} \right) \qquad (20)$$

where

$$\Delta_k = \sum_q \frac{|B_{k,q}|^2}{\hbar \omega_q} \qquad (21)$$

4. The Calculation of Critical Temperature Tc

Bipolaron is a boson. Under some condition in the system consisting of bipolaron the condensation phenomena will occur. The system will be in the superconductivity state. In the following we calculate the critical temperature Tc using grand canonical ensemble.

We can change the form of the Hamiltonian of the system (20) as follows

$$H = \sum_k E_k n_k + \sum_q E_q \left( n_q + \frac{1}{2} \right) \qquad (22)$$

where

$$E_k = E_0(k) - \Delta_k, \qquad E_q = \hbar \omega_q$$

$$n_k = B_k^+ B_k, \qquad n_q = a_q^+ a_q$$

We will make a calculation of critical temperature Tc for the system. The strategy is that we regard the system as a system of the gas of boson particles. They obey the boson statistical law within grand canonical ensemble. Firstly, we calculate the grand partition function Zg. According to the definition of grand partition function [10]

$$Z_g = Tr \exp\left[ -\beta (H - \mu N) \right] \qquad (23)$$

where

$H$ is the Hamiltonian of boson particle gas. It has the form as

Eq. (22).

$N$ is the operator of the total number of particle of the system.

$\mu$ is the chemical potential.

and $\beta = \dfrac{1}{k_B T}$.

Secondly, we can calculate the thermodynamically potential $\Omega$. Here,

$$\Omega = -k_B T \ln(Z_g) \quad (24)$$

and according to the definition we can calculate the total number of particle

$$N = \frac{\partial \Omega}{\partial \mu} \quad (25)$$

Thirdly, we write down the Helmholtz free energy,

$$F = \Omega + \mu N \quad (26)$$

Using the stability condition of system of boson ( bipolaron )

$$\partial F = \frac{\partial F}{\partial \Delta} \partial \Delta \quad (27)$$

We will obtain the satisfying equation of Tc.

In the occupation representation the state vector of boson particle system is

$$|\{n\}\rangle = |\{n_k\},\{n_q\}\rangle = |\{n_k\}\rangle|\{n_q\}\rangle$$

$$|\{n_{k1}\}\rangle|\{n_{k2}\}\rangle\cdots|\{n_{ki}\}\rangle\cdots|\{n_{q1}\}\rangle|\{n_{q2}\}\rangle\cdots|\{n_{qi}\}\rangle\cdots \quad (28)$$

According to the definition of the grand partition function

$$Z_g = Tr \exp[-\beta(H - \mu N)] = \sum_{\{n\}} \langle\{n\}|\exp[-\beta(H - \mu N)]|\{n\}\rangle \quad (29)$$

In the occupation representation, the operator in above equation can be replaced by their eigenvalue

$$Z_g = \sum_{\{n\}} \langle \{n\} | \exp\left(-\beta \left[\sum_k E_k N_k + \sum_q E_q \left(N_q + \frac{1}{2}\right) - \mu N\right]\right) | \{n\} \rangle$$

$$= \sum_{\{n_k\}} \langle \{n_k\} | \exp\left(-\beta \left[\sum_k (E_k - \mu) N_k\right]\right) | \{n_k\} \rangle \times$$

$$\times \sum_{\{n_q\}} \langle \{n_q\} | \exp\left(-\beta \left[\sum_q (E_q - \mu) N_q + \frac{E_q}{2}\right]\right) | \{n_q\} \rangle$$

$$= \sum_{n_{k1}} \langle n_{k1} | \exp\left(-\beta \left[\sum_{k1} (E_{k1} - \mu) N_{k1}\right]\right) | n_{k1} \rangle \sum_{n_{k2}} \cdots \sum_{n_{ki}} \cdots \times$$

$$\times \sum_{n_{q1}} \langle n_{q1} | \exp\left(-\beta \left[\sum_{q1} (E_{q1} - \mu) N_{q1} + \frac{E_{q1}}{2}\right]\right) | n_{q1} \rangle \sum_{n_{q2}} \cdots \sum_{n_{qi}} \cdots$$

$$= \prod_k \frac{1}{1 - \exp(-\beta[\xi - \Delta])} \prod_q \frac{1}{1 - \exp(-\beta E_q)} \tag{30}$$

where

$$\xi = E_0(k) - \mu$$

regardless the zero point vibrational energy $\frac{E_g}{2}$ and $\mu = 0$ for phonon.

Therefore, from equations (24), (25), (26), (27), (28) and (30), we have

$$\mu \beta \exp(-\beta[\xi - \Delta]) = \exp(-\beta[\xi - \Delta]) - 1 \tag{31}$$

In the critical state $\Delta = 0$, and T= Tc we obtain the equation

$$\mu \beta_c \exp(-\beta_c \xi) = \exp(\beta_c \xi) - 1 \tag{32}$$

$$\beta_c = \frac{1}{k_B T_c}$$

Equation (32) is the transcendental equation. It determines the Tc.

Solving equation ( 32 ) we can get

$$T_c = \frac{\mu}{k_B} \quad (33)$$

After considering the modification of $\mu$ with temperature, we have

$$k_B T_c = 3.31 \frac{\hbar^2}{m^{**}} n^{2/3} \quad (34)$$

where $m^{**}$ is the mass of bipolaron and $n$ is the concentration of bipolaron.

From the data of Savransky S.D.[11,12], we can explain the experimental results of I.V. Berman and N.B. Brandt.[2A]. If we take $n = 8 \times 10^{18} cm^{-3}$ [11] and $m^{**} = 100 m_e$ [12]. We can get Tc=1.2 K. This is the result of Fig.14 of ref. 2A).

## 5. Discussion

From this paper and other papers [13,14], we can see that the mathematical method and physical model we have developed from the investigation of high Tc superconductivity also can use in room temperature superconductivity and low temperature superconductivity in one dimension model. We can say if people want to seek high temperature materials we must sort the one dimensional materials like Oxidized Polypropylene. From our further research [13,14] we can see that this mathematical method and physical model we have developed can use in other field of physics, for example, the superconductivity in the core of Neutron Star and the prediction of possibility of superfluidity of Optical Soliton in some materials and some environment ( temperature ). We expect, our method and model developed here will encourage more applications in the future.